          [2004/06/22 1.2 (KOF); 2001/04/25 1.1 (PWD)]

\documentclass[a4paper,twocolumn]{Gaia2004} % European paper size
\usepackage{times}      % for font
\usepackage{epsfig}     % for figure inclusion
\usepackage{natbib}     % for bibliography
\title{Probing Galactic reddening with the 8620$\,$\AA$\ $diffuse interstellar band}

\author{S.~Vidrih, }
\author{T.~Zwitter}
\affil{Faculty of Mathematics and Physics of University of Ljubljana, Jadranska 19, 1000 Ljubljana, Slovenia}

\bibpunct{(}{)}{;}{a}{}{,}  % to set bibliography punctuation to A&A style

\begin{document}


\maketitle

\begin{abstract}
  
Using the correlation between the equivalent width of the diffuse interstellar band (DIB) at 8620\,\AA\ and the interstellar reddening reported by~\citet{mun00} GAIA could directly trace the interstellar extinction throughout the Galaxy. We checked for the magnitude and distance limitations of this method in 42 Galactic directions by simulating the RVS data on the stellar sample provided by the model of stellar population synthesis of the Galaxy~\citep{rob03}. The simulation indicates that the imprint of the 8620\,\AA\ DIB will be detected in the RVS spectra of stars with magnitudes up to $V\sim16$ with sufficient accuracy to trace not only the distribution of the interstellar medium but also the radial component of its kinematic motion, i.e. the radial velocity of the mass center of the dust cloud in the line of sight.

\end{abstract}

\section{Introduction}

A surprisingly good correlation between the equivalent width ($EW$) of the diffuse interstellar band (DIB) at 8620\,\AA\ and the interstellar reddening ($E_{B-V}$) has been reported by~\citet{mun00}. The relation was calibrated as $E_{B-V}=2.69\times EW(\mathrm{\AA})$ on the base of 37 stars widely distributed in Galactic coordinates and distances. However, the exact slope of the relation is expected to depend on the properties of the interstellar medium and to vary with direction and distance. The most general relation would thus be
\begin{displaymath}
E_{B-V}=\alpha(l,b,D)\times EW(\mathrm{\AA})\quad .
\end{displaymath}

GAIA will determine $E_{B-V}$ from the photometric measurements. If the 8620\,\AA\ DIB will be resolved from RVS spectra, the fine-grid calibration of $\alpha(l,b,D)$ will be performed and thus the distribution and the chemical composition of the interstellar matter will be obtained. Moreover, the exact wavelength position of the 8620\,\AA\ DIB will reveal the information on the kinematics of the interstellar matter.

The purpose of the here presented simulation was to check for the limitations and to estimate the distance range of this interstellar mapping method.

\section{Simulation}

\begin{figure}[h]
  \begin{center}
    \leavevmode
 \includegraphics[clip,width=1\linewidth]{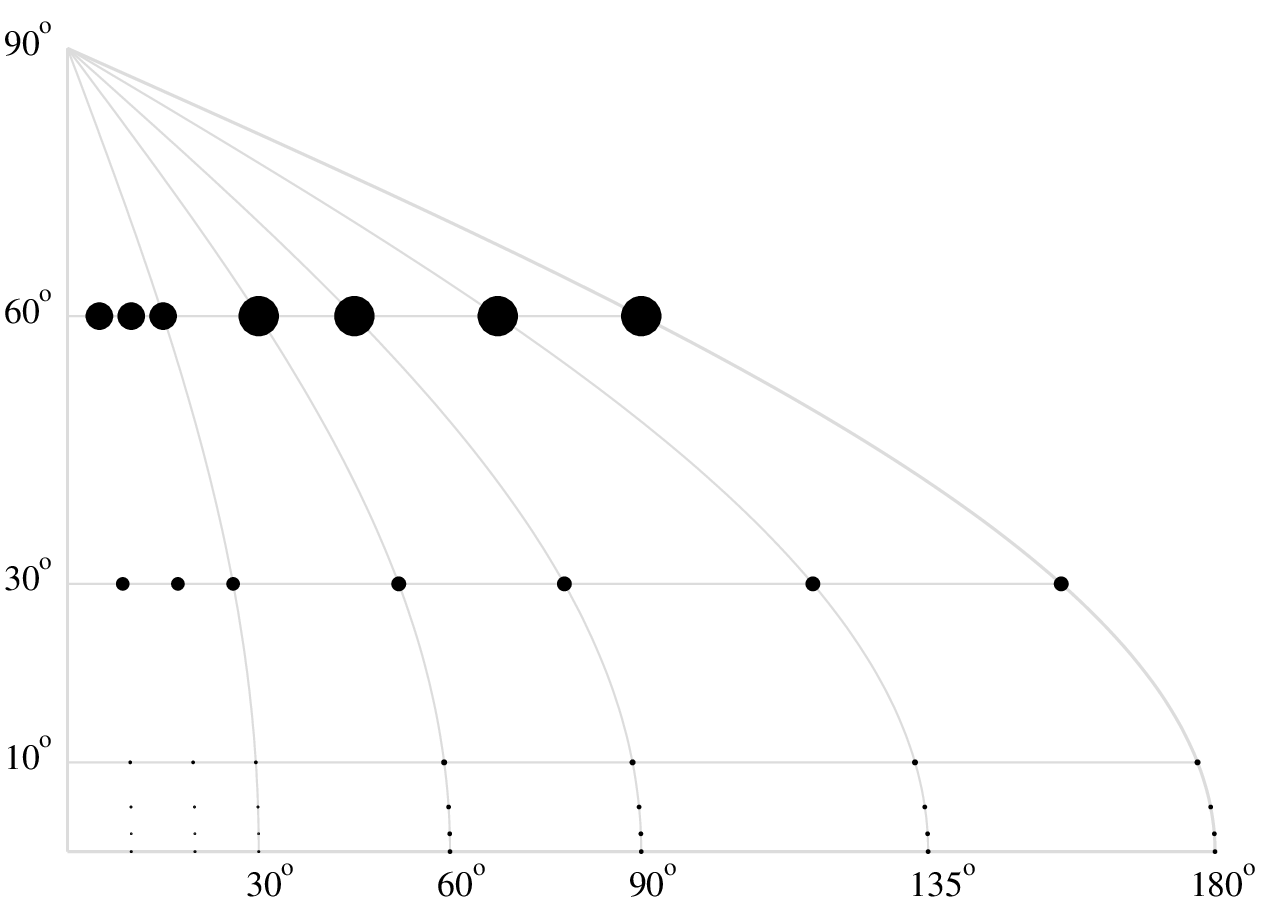}
   \end{center}
  \caption{The Galactic stellar population was simulated in 42 directions. The solid angle of each simulation was chosen in such a way that the number of simulated stars was kept approximately constant, i.e. $\sim5 000-15000$. The sizes of the solid angles that varied between 0.2 and 15 square degrees are presented in a relative scale with the sizes of the black dots.}
  \label{galactic_results_angle}
 \end{figure}

\begin{figure*}[t]
  \begin{center}
    \leavevmode
 \includegraphics[clip,width=0.7\linewidth]{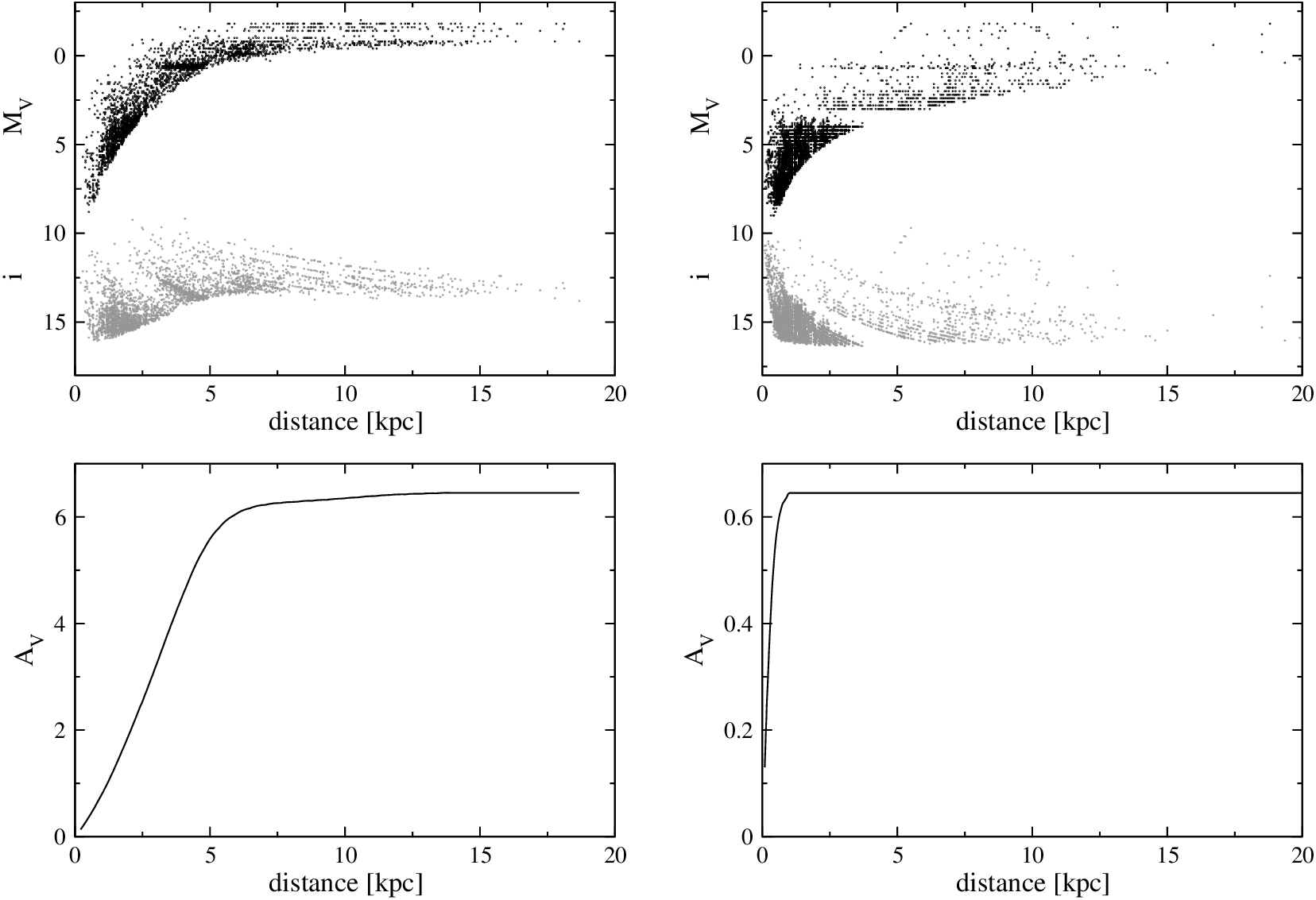}
   \end{center}
  \caption{Two exemplary simulated stellar samples in the directions $l=10^\circ$, $b=2^\circ$ (left) and $l=10^\circ$, $b=30^\circ$ (right). Above: Absolute $V$ magnitudes (black) and apparent $I$ magnitudes (gray) of the simulated stars as a function of distance. Below: Interstellar extinction according to~\citet{ds01} as a function of distance in these two Galactic directions.}
  \label{stellar_sample}
 \end{figure*}

The Galactic stellar population was simulated with the Besan\c con model of stellar population synthesis~\citep{rob03}. The interstellar extinction was taken into account with the use of the three-dimensional model of the dust distribution by~\citet{ds01} and~\citet{dri03}. Simulations were performed in 42 Galactic directions ($l=10^\circ$, $l=20^\circ$, $l=30^\circ$, $l=60^\circ$, $l=90^\circ$, $l=135^\circ$ and $l=180^\circ$; $b=0^\circ$, $b=2^\circ$, $b=5^\circ$, $b=10^\circ$, $b=30^\circ$ and $b=60^\circ$) with the solid angles varying between 0.2 and 15 square degrees for different Galactic directions in order to keep the number of the simulated stars approximately constant, i.e. between 5000 and 15000. Simulated Galactic directions are shown in Figure~\ref{galactic_results_angle}. Two exemplary stellar samples are presented in Figure~\ref{stellar_sample}.

The simulation of the DIB detection accuracy was performed as follows. For any star given by the Besan\c con model the closest computed synthetic spectrum~\citep{zwi04} from the multi-parameter spectral grid was taken. The spectrum was then shifted for the star's radial velocity and the 8620\,\AA\ DIB Gaussian line, Doppler shifted for between 0 and the shift of the star was added. The width of the Gaussian line was taken as in~\citet{mun00}, i.e. $FWHM$ was taken to be $5.5\,\mathrm{\AA}$ and the amplitude was determined from $EW$. Finally, the noise appropriate to the star's magnitude and the RVS instrument~\citep{mun03} was added to the synthetic spectrum. In addition to the simulated star spectrum the synthetic spectrum template was generated, which differed from the simulated spectrum for one step in each of the parameters (temperature, metallicity\ldots) of the spectral grid. It also differed in the radial velocity~\citep{kat04} and possessed no DIB line. The generated template was subtracted from the RVS simulated spectrum and the 8620\,\AA\ DIB line was resolved from the residuum by fitting the Gaussian line with three parameters, i.e. $EW$, $\sigma$ and $\lambda_0$ on it.

\section{Results}

\begin{figure}[!h]
  \begin{center}
    \leavevmode
 \includegraphics[clip,width=1\linewidth]{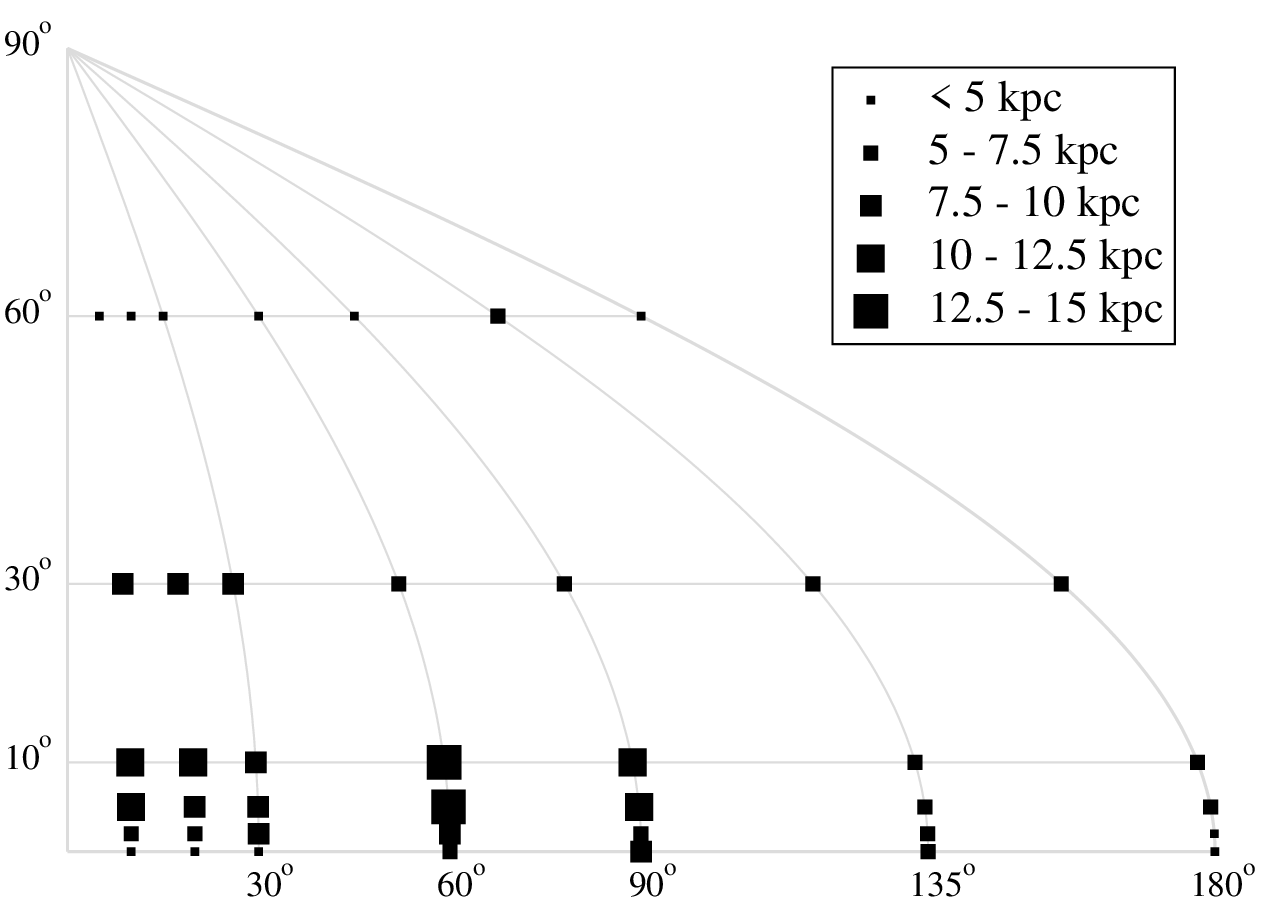}
   \end{center}
  \caption{The Galactic map with achieved distance ranges for the interstellar matter mapping via the 8620\,\AA\ DIB line. This method of the interstellar mapping is the most efficient at intermediate latitudes, where the interstellar extinction is already moderate but still high enough to leave the significant imprint in the stellar spectra.}
  \label{galactic_results_distance}
 \end{figure}

\begin{figure*}[!p]
  \begin{center}
    \leavevmode
 \includegraphics[clip,width=0.7\linewidth]{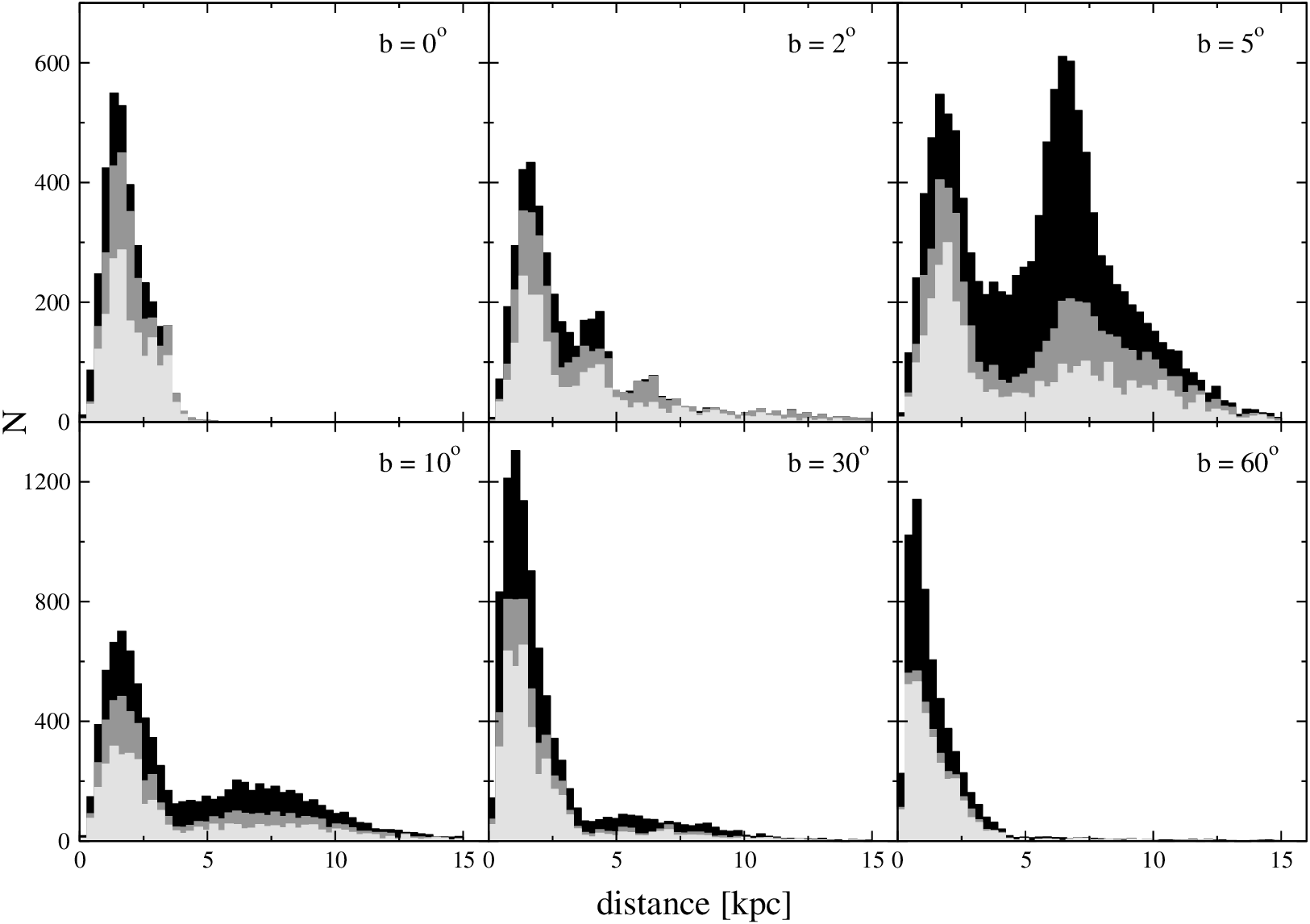}
   \end{center}
  \caption{Recovery of the 8620\,\AA\ DIB line from the simulated RVS spectra: Number of simulated stars as a function of distance in six Galactic directions ($l = 10^\circ$; $b = 0^\circ$, $b = 2^\circ$, $b = 5^\circ$, $b = 10^\circ$, $b = 30^\circ$ and $b = 60^\circ$). Black color represents all simulated stars, dark gray those for which the DIB line was recovered with an accuracy of at least $30\,\mathrm{km\,s^{-1}}$ and light gray those for which the recovery of the DIB line was more accurate than $1\,\mathrm{km\,s^{-1}}$.}
  \label{l010_distance_histogram}
 \end{figure*}

\begin{figure*}[!p]
  \begin{center}
    \leavevmode
 \includegraphics[clip,width=0.7\linewidth]{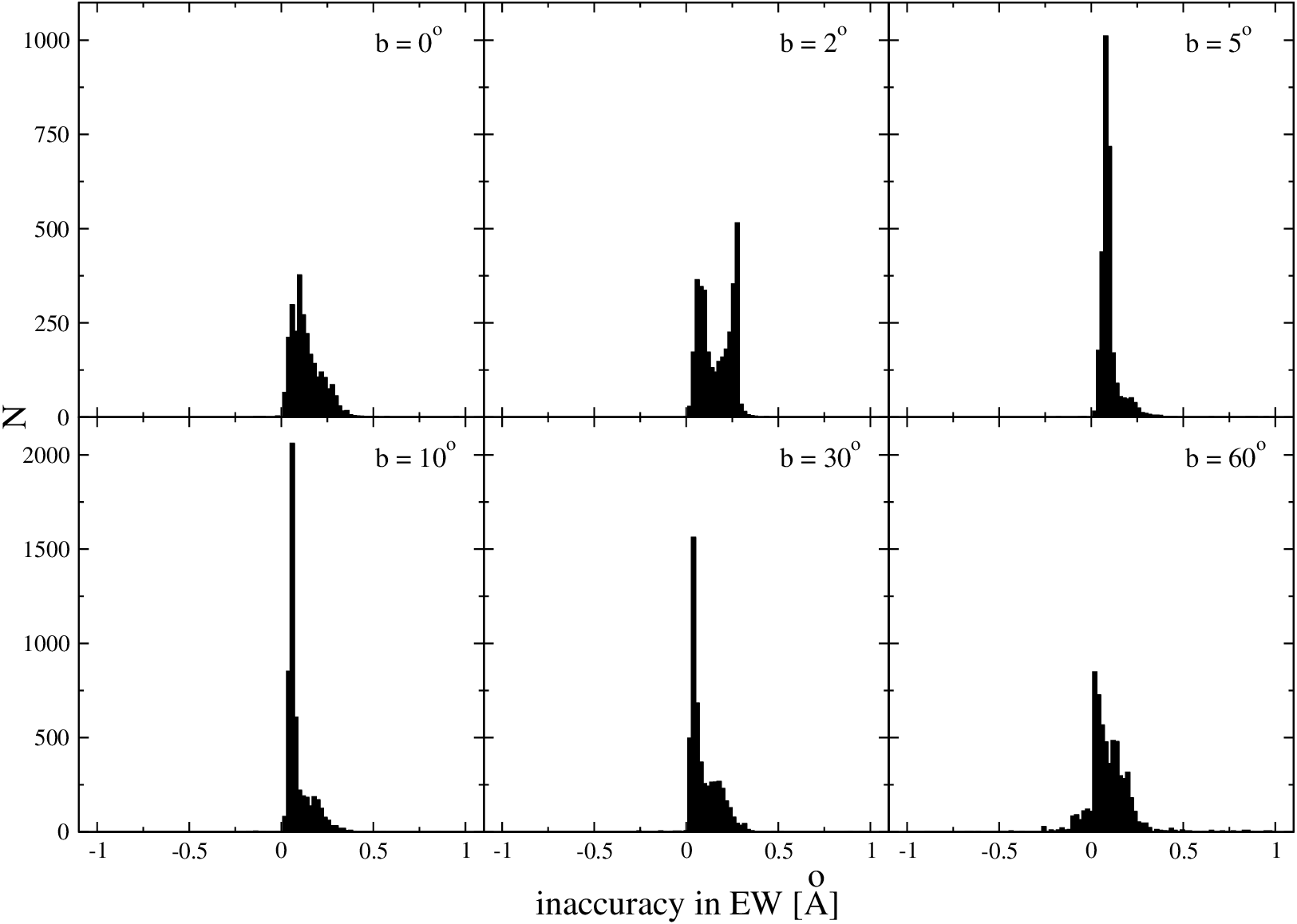}
   \end{center}
  \caption{Recovery of the 8620\,\AA\ DIB line from the simulated RVS spectra: Number of simulated stars as a function of inaccuracy in equivalent width, i.e. the difference between the simulated and recovered $EW$ of the DIB line. According to~\citet{mun00} $EW$ is proportional to $E_{B-V}$, where $EW = 0.5\,\mathrm{\AA}$ corresponds to $E_{B-V} = 1.35$.}
  \label{l010_ew_diff_histogram}
 \end{figure*}

Interstellar extinction is responsible for both, the imprint of the 8620\,\AA\ DIB line in the stellar spectra and the increase of stellar magnitudes that limits the detection of the DIB line. It is not surprising, that the simulation results indicate the following. High interstellar extinction is a strong limiting factor in the Galactic plane where the achievable range of the interstellar matter mapping via the 8620\,\AA\ DIB line is smaller than 5\,kpc. When going out from the Galactic plane the 8620\,\AA\ DIB line can be detected in stellar spectra up to distances 10-15\,kpc. Finally, at very high latitudes the imprint of small interstellar extinction becomes weak, which again reduces the detection of the DIB line. Tracing of the 8620\,\AA\ DIB line varies also with the Galactic longitude. The Galactic map with the achieved distance ranges for the interstellar matter mapping via the 8620\,\AA\ DIB line is shown in Figure~\ref{galactic_results_distance}. Results of the 8620\,\AA\ DIB line recovery from the simulated RVS spectra in six Galactic directions are shown in Figures~\ref{l010_distance_histogram} and~\ref{l010_ew_diff_histogram}. 

\clearpage

To conclude, simulated RVS data indicate that the 8620\,\AA\ DIB line will be detected in the RVS spectra of relatively faint stars with magnitudes up to $V\sim~16$ with sufficient accuracy to trace not only the distribution of the interstellar medium but also the radial component of its kinematic motion. However, it is necessary to point out that the obtained results of the simulation should be further constrained by more careful treatment of the S/N issue. Other realistic problems such as crowding in the stellar field should also be taken into account.

\section*{Acknowledgments}
We thank Annie Robin from the Observatoire de Besan\c con for generating the Galactic stellar population which was used as the input for our simulation. The financial support from the conference organizers is kindly acknowledged.

\end{document}